\documentstyle[11pt,newpasp,twoside,epsf]{article}

\begin{document}

\title{Massive Stellar Clusters in Interacting Galaxies}

\author{Bryan W. Miller}
\affil{Sterrewacht Leiden, Postbus 9513, 2300 RA Leiden, NL}

\begin{abstract}
Massive clusters are now seen to form easily in interacting and
merging galaxies, making these excellent environments for studying the
properties of young clusters.  New observations of the Antennae
(NGC~4038/39) show that the most luminous young clusters do not have a
measurable tidal radius.  Most observations suggest that the
luminosity function (LF) and mass functions of young clusters are
single power laws.  However, there are many uncertainties at the faint
end of the LF.  For example, contamination from massive stars may be
important.  The shape and evolution of the LF, and more fundamentally,
the mass function, of massive clusters had implications for our
understanding of both the formation and the destruction of massive
stellar clusters.
\end{abstract}

\section{Introduction}

Ten years ago there were only suggestions that interacting and merging
galaxies contained young, massive star clusters (Schweizer 1982; Lutz
1991).  However, subsequent observations, especially those using the
{\it Hubble Space Telescope}, have shown that young clusters are
nearly ubiquitous in such systems.  Table~1 of Schweizer (1999) gives
a nearly complete list of galaxies observed to have young star
clusters.  To this list can be added recent papers on NGC~3597
(Carlson et al.\ 1999; Forbes \& Hau 1999) and NGC~5128 (Holland et
al.\ 1999).  It would seem that cluster
formation is a natural result of the star formation triggered by strong
gravitational interactions or direct collisions.  

Many of these observations were motivated by the question of whether
ellipticals can form from the mergers of two spirals, but, in
addition, they provide important information about the formation and
evolution of the star clusters themselves.  The sizes and profiles of
the youngest clusters can constrain their initial states.  The
distribution of ages gives the cluster formation history, which can be
compared with the dynamical and star formation histories.  The ages
and metallicities allow us to determine masses and the mass
function. This is critical for understanding the physics of how
clusters form.  The evolution of the mass function then shows us how
the interplay between stellar evolution and both internal and external
dynamics affect cluster evolution.  This paper will review the sizes,
ages, and masses of young star clusters in merging galaxies.  The
focus will be on recent results from WFPC2 observations of NGC~4038/39
(``the Antennae'', Whitmore et al.\ 1999, hereafter W+99; Zhang \&
Fall 1999, hereafter ZF99) and NGC~7252 (Miller et al.\ 1997,
hereafter M+97).

\section{Cluster Sizes}

The sizes of the star cluster candidates determine whether they are
structurally similar to Galactic globular clusters (GCs) with
effective radii of a few parsecs, or to open clusters and associations
which can have a much wider range of sizes but which are generally
bigger than GCs.  The sizes and profiles of the youngest clusters are
also important initial conditions for dynamical models of clusters
(see Zwart's comments in the Discussion).  In addition, Galactic GCs
have a lognormal or broken power-law mass function with a
characteristic mass of about $10^5$~M$_{\sun}$, while open clusters
have an unbroken power-law mass function.  Thus, density could be
related to formation process.  Pre-refurbishment {\it HST}\ images of
NGC~4038/39 and NGC~7252 showed the cluster candidates to have $R_{\rm
eff} \ga 10$~pc and a power-law luminosity function.  Thus, it was
argued that these objects would not become GCs and that galaxy mergers
would not produce GC systems like seen in elliptical, calling into
question whether ellipticals were produced by mergers (van den Bergh
1995)

Observations with the corrected optics of WFPC2 have consistently
shown that the bulk of the young cluster candidates are marginally
resolved and that $R_{\rm eff} \la 5$ pc (Schweizer et al.\ 1996;
Whitmore et al.\ 1997; M+97; Carlson et al.\ 1999;
W+99).  Thus, the effective radii are consistent with the values for
old GCs in M87 (Whitmore et al.\ 1995) and the Milky Way.  As
suspected, the larger effective radii measured previously were due the
difficulty of measuring sizes on the aberrated WF/PC1 images.

In the Antennae we may now be seen changes in both the effective and
tidal radii with age (see Section~3).  Old cluster candidates have
$\langle R_{\rm eff} \rangle = 3.0 \pm 0.3$~pc while young and
intermediate age cluster candidates have $\langle R_{\rm eff} \rangle
= 4.6 \pm 0.4$~pc.  Further, the tidal radii of the young clusters can be
much larger than for the old clusters (Figure~1).  Thus, the density
distribution of clusters may extend beyond their tidal radii at birth
and a few orbits around the galaxy are needed to remove the stars
beyond the tidal radius.

\begin{figure}
\plotfiddle{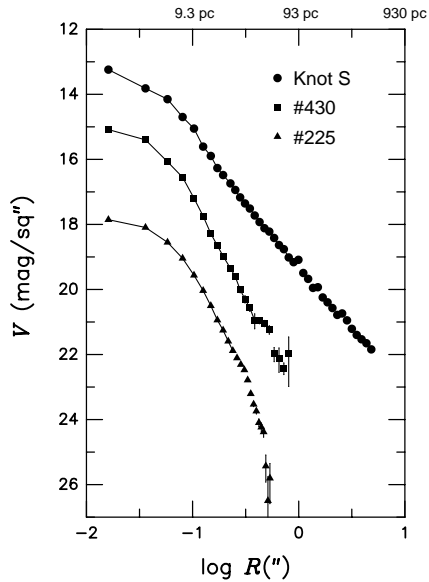}{7cm}{270}{40}{40}{-120}{260}
\caption{Radial surface brightness profiles of three star clusters in
NGC~4038/39: the very extended and luminous Knot~S, its neighbor \#430
(both with ages $<10$~Myr), and the 500~Myr-old cluster \#225.
Neither of the youngest clusters have a measurable tidal limit, unlike
the older cluster which has a tidal radius of about 50 parsecs
(W+99).}
\end{figure}

\section{Metallicities and Ages}

Since broad-band colors are degenerate in age and metallicity, we must
have an independent measurement of one of these properties in order to
determine the other from evolutionary models.  Spectroscopy of the
brightest three young clusters in NGC~7252 shows that they have
near-solar metallicity (Schweizer \& Seitzer 1998), so solar
metallicity is assumed for all the young clusters.  Then, ages are
determined by comparing the broad-band colors and luminosities with
evolutionary models for simple stellar populations.  The youngest
clusters are often surrounded by considerable dust, so we attempt to
correct for this internal extinction by calculating ``reddening-free''
indices based on the (\ub), (\bv), and ($V\!-\!I$) colors (M+97; W+99).  In
NGC~4038/39, the youngest clusters are also be distinguished by their
H$\alpha$ emission.

Multiple populations of star clusters can be distinguished  in several
systems.  Four populations have been identified in the Antennae: 1) a
$<10$~Myr-old population with compact H$\alpha$ emission located near
the dusty overlap region; 2) a $\sim100$~Myr-old population found
further out in the disk of NGC~4038; 3) a $\sim500$~Myr-old
population that may have been formed during the first close encounter
when the tidal tails were formed; and 4) a few $\sim10$~Gyr-old
clusters that are probably original GCs from the progenitor galaxies
(W+99).  The very young ages of the youngest clusters are confirmed by
ultraviolet spectroscopy (W+99) and infrared spectroscopy (see the
contributions by Gilbert and Mengel).
The older merger remnant NGC~7252 has a $<10$~Myr-old population of rather
extended clusters associated with the central gas disk, a
500--800~Myr population that formed during the merger, and old
clusters from the progenitor galaxies (M+97).  Young
cluster formation lasts for several hundred Myr, consistent with the
dynamical time-scale of the merger event.

\section{Luminosity and Mass Functions}

WFPC2 observations of young star clusters have most often found the
luminosity functions (LFs) to have a power-law shape, $\phi(L) \propto
L^{\alpha}$, with $\alpha\approx -1.8$ down to the completeness limits
of the observations (e.g. Schweizer et al.\ 1996; M+97; Carlson et
al.\ 1999).  The masses of the most luminous clusters, as inferred
from evolutionary models of the appropriate age and metallicty, can
approach $10^8$~M$_{\sun}$, over an order of magnitude more massive
than the most massive Galactic GC (Schweizer \& Seitzer 1998).  These
are extreme clusters, even considering the fact that the mass-to-light
ratios of the models are about a factor of two higher than measured
(Fritze-v.~Alvensleben, this proceedings).  At the faint end, the new
observations are sensitive enough detect objects less massive than
$10^5$~M$_{\sun}$, the mass at the peak of the old GC mass function.
If the mass function were peaked, then one would expect to see a bump
in the luminosity function since fading preserves the shape of the
luminosity function.  This immediately suggests that the mass function
is a power law.  However, one must be sure that large relative age
spreads, reddening, and and stellar contamination do not affect the
shape of the cluster luminosity function (cf. Meurer 1995).

\begin{figure}
\plotfiddle{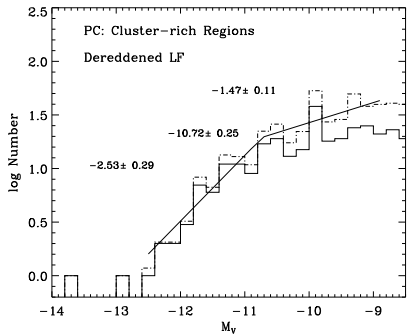}{4cm}{0}{30}{30}{-80}{-30}
\caption{Reddening and completeness-corrected LF of young star
clusters in NGC~4038/39.  The cluster candidates are taken from
regions on the PC which are dominated by clusters rather than stars
(W+99). The LF has a bend at a luminosity corresponding to
$\sim10^{5}$~M$_{\sun}$.}
\end{figure}  

A few observations are now suggesting that the young cluster LF may not
be a single power law.  Zepf et al.\ (1999) find that the LF for young
clusters in NGC~3256 to be slightly flattened for $M_B > -11$.
However, the statistical significance of the flattening is relatively
weak (2.5$\sigma$) and the most likely mass function is a power law
with $\alpha=-1.8$.  The new observations of the Antennae show a
stronger flattening of the luminosity function (Figure~2).  The break
occurrs at a mass of $\sim10^5$~M$_{\sun}$, similar to the peak in the
old GC mass function.  While this is suggestive that the mass function
has a break or peak, a reconstruction of the mass function by ZF99
shows that it is still most likely a single power law with
$\phi(M) \propto M^{-2}$.

The proximity of the Antennae, the depth of the photometry, and the
youth of the starburst made stellar contamination a significant issue.
However, stellar contamination at the faint end of the cluster LF may
be significant even for the older and more distant merger remnants like
NGC~3921 and NGC~7252.

I have attempted to determine the young cluster mass function in
NGC~7252 by matching the observed LF with Monte Carlo simulations.
Artificial clusters are drawn from either a power-law mass function
consistent with the GC mass function of the Galaxy, or a power-law
mass function with slope equal to the observed slope of the cluster
LF.  The masses are converted to magnitudes and colors using Bruzual
\& Charlot (1996) evolutionary models of simple stellar populations.
Young clusters are assumed to have solar metallicty, and the mean age
is determined by matching the ($V\!-\!I$) colors of the clusters.  Some
simulations also include artificial stars drawn from a
Salpeter IMF and placed in the color-magnitude diagram using Geneva
evolutionary tracks (Schaller et al.\ 1992) and
bolometric corrections from Bessell, Castelli, \& Plez (1998) for 
solar metallicity.

Measurement error and selection criteria are applied to the
simulations before comparing them with the observations.
Gaussian-distributed random errors are added to the colors and
magnitudes of model clusters and stars based on the photometric
uncertainties of the observed clusters.  Then, both the observed and
simulated clusters are selected according to the same criteria. The
goodness-of-fit between a simulation and the observations is measured
by the $\chi^2$ per degree of freedom, $\chi^2/\nu$.

\begin{figure}
\plotfiddle{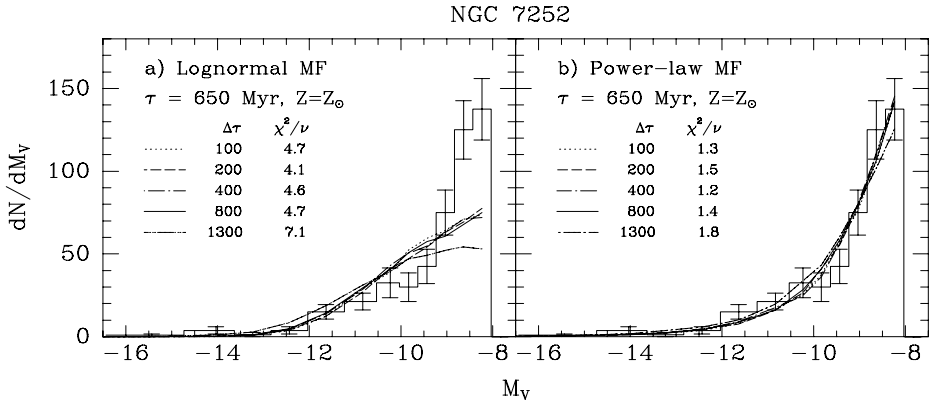}{5cm}{90}{60}{60}{230}{-65}
\caption{Luminosity functions for observed and
simulated star clusters in NGC~7252.  The mean age of the simulated
clusters is $\tau = 650$~Myr and the curves are for different age
spreads, $\Delta\tau$ [Myr], which do not have a significant effect.  a) The
mass function is lognormal.  These models are not good matches to the
data. b) The mass function is a power-law with $\alpha=-1.8$.  Values
of $\chi^2/\nu \approx 1$ indicate that the simulations are good
representations of the observations.}
\end{figure}

If all the observed objects are clusters, then the mass function is a
power law with slope $\alpha \approx -1.8$ (Figure~3).  However,
stellar contamination at the faint end of the cluster LF may be 
important.  Using a star formation rate (SFR) measured from
H$\alpha$ images results in more objects at faint magnitudes than are
observed.  With young cluster drawn from a lognormal mass function,
the SFR needed to match the observed LF is about half the observed SFR
(Figure~4).  This difference could be explained if the binary fraction
is about 50\%.  The observed SFR could be much lower if the H$\alpha$
flux is due to shocks or other processes besides star formation.  Most
of the flux in the region under consideration is from diffuse emission
rather than discreet HII regions.  The mass-loss prescription in the
stellar evolutionary tracks is also a crucial parameter; higher
mass-loss rates yield fewer supergiants. The main point is that
determining cluster mass functions is complicated, all these factors
must be considered.

\begin{figure}
\plotfiddle{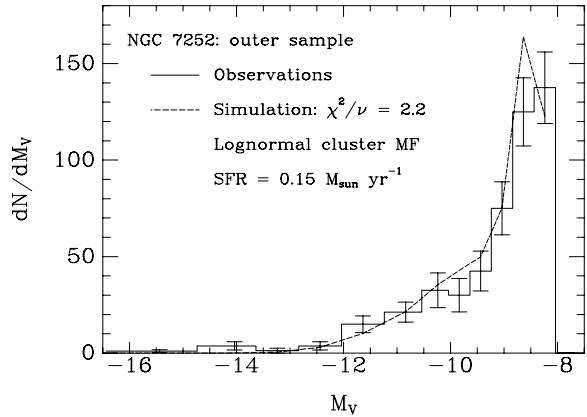}{5cm}{90}{35}{35}{150}{-40}
\caption{As Figure~3, except that the mass function of the simulated
clusters is lognormal and the steeply rising faint end of the LF is
populated by supergiant stars formed with ${\rm SFR} = 0.15$ M$_{\sun}$
yr$^{-1}$.  The statistical match to the data is similar to the
power-law models in Figure~3b.}
\end{figure}

\section{Conclusions}

The study of young star clusters in interacting and merging galaxies
is a good example of the symbiotic relationship between observations
and theory.  Observations of the youngest star clusters in the
Antennae are providing the initial density distributions and the
initial cluster mass function that are needed for models of individual
star clusters and cluster systems.  Further observations of older
merger remnants will hopefully show how the mass function evolves with
time.  On the other hand, evolutionary models are needed to convert
colors and luminosities into ages and masses.  Dynamical models will
explain the processes that cause clusters and cluster systems to evolve.

The current observations suggest that young clusters may be born
without a tidal radius, that cluster formation occurs over several
hundred Myr in a merger, and that the mass function of young clusters
is most likely a power-law (though there are indications of
flattening).  Thus, there still appears to be a difference between the
mass functions of young clusters and old GCs.  This could be due
either to the effect of different initial conditions at recent epochs
(e.g. increased metallicity) or to the slow destruction of low-mass
clusters over a Hubble time (see ZF99 and references therein).

\acknowledgments

I would like to thank the Leids Kerkhoven Bosscha Fonds for a subsidy
that allowed me to attend this workshop.  
Rob Kennicutt and Audra Baleisis kindly provided the H$\alpha$ image
of NGC~7252.  Thanks to Michael Fall, Brad Whitmore, and
Gerhardt Meurer for many useful suggestions on modeling the cluster mass
function.

\section*{Discussion}

\noindent
{\bf C. Boily:} How is the tidal radii of candidate clusters
estimated?  What might be the velocity dispersion of populations of
candidate clusters?\\
{\bf B. Miller:} The tidal radius depends on the shape of the
potential, which is complicated in a merging system.  However, the
potential of a young merger like NGC~4038 may still not be too different
from a normal disk.  Thus, we can get a range of likely tidal radii by
looking at the Galaxy, M31, and truncated clusters in the Antennae
itself.  They do seem to be similar, with values of $r_t = 50-100$~pc.  
The velocity dispersion of the cluster system is probably on
the order of 100 km sec$^{-1}$ (see Schweizer \& Seitzer 1998).

\noindent
{\bf S. P. Zwart:} The luminosity density of your youngest
globular cluster seems to extend beyond its tidal radius in the
potential of its parent galaxy.  the small age of this system may be
smaller than the cluster's crossing time, which suggests that when
clusters form their density distribution extends beyond the tidal
radius.  This is important for understanding the initial conditions of
globular clusters, which are required for numerical models.

\noindent
{\bf P. Kroupa:} The flattening of the LF near 10$^{5}$ M$_{\sun}$
may be due to cluster-cluster disruptions in cluster-rich regions.
Are there any {\it HST} images for tidal dwarf galaxies?\\
{\bf B. Miller:}  {\it HST} images of tidal dwarf candidates in 
NGC~4038/39 and NGC~7252 have been taken and are being analyzed.

\noindent
{\bf J. C. Mermilliod:} The important point is to distinguish
a real bound star cluster from a large OB association which can cover
several hundred parsecs.  Galactic examples would be the Sco OB1
region with a dense cluster and a whole population of supergiants, or
the h and $\chi$ Persei region.\\
{\bf B. Miller:} Agreed, we try to select as cluster candidates only
the most compact objects that do not appear to be stars.

\end{document}